\documentclass[10pt, conference]{IEEEtran}
\usepackage{cite}
\usepackage{amsmath,amssymb,amsfonts}
\usepackage{algorithmic}
\usepackage{textcomp}

\usepackage{url}
\usepackage{listings}
\usepackage{multicol}
\usepackage{enumitem}
\usepackage[pdftex]{graphicx,xcolor}

\DeclareMathOperator{\argmax}{argmax}

\begin{document}

\setlength{\floatsep}{1mm}
\setlength{\textfloatsep}{1mm}

\title{Improving Semantic Consistency of Variable Names with Use-Flow Graph Analysis}

\author{
\IEEEauthorblockN{1\textsuperscript{st} Yusuke Shinyama}
\IEEEauthorblockA{\textit{dept. name of organization (of Aff.)} \\
\textit{Tokyo Institute of Technology}\\
Meguro-ku, Tokyo, Japan \\
euske@sde.cs.titech.ac.jp} \vspace*{-1em}
\and
\IEEEauthorblockN{2\textsuperscript{nd} Yoshitaka Arahori}
\IEEEauthorblockA{\textit{dept. name of organization (of Aff.)} \\
\textit{Tokyo Institute of Technology}\\
Meguro-ku, Tokyo, Japan \\
arahori@c.titech.ac.jp} \vspace*{-1em}
\and
\IEEEauthorblockN{3\textsuperscript{rd} Katsuhiko Gondow}
\IEEEauthorblockA{\textit{dept. name of organization (of Aff.)} \\
\textit{Tokyo Institute of Technology}\\
Meguro-ku, Tokyo, Japan \\
gondow@cs.titech.ac.jp} \vspace*{-1em}
}

\maketitle

\begin{abstract}
Consistency is one of the keys to maintainable source code and hence a
successful software project.  We propose a novel method of extracting the
intent of programmers from source code of a large project ($\sim 300$ kLOC)
and checking the semantic consistency of its variable names.
Our system learns a project-specific naming convention for variables
based on its role solely from source code, and suggest alternatives
when it violates its internal consistency. The system can also show
the reasoning why a certain variable should be named in a specific
way. The system does not rely on any external knowledge. We applied
our method to 12 open-source projects and evaluated its results with
human reviewers. Our system proposed alternative variable names for
416 out of 1080 (39\%) instances that are considered better than ones
originally used by the developers.  Based on the results, we created
patches to correct the inconsistent names and sent them to its
developers. Three open-source projects adopted it.
\end{abstract}



\section{Backgrounds}



A large software project is typically developed and maintained by a
number of people. Even if the project is relatively small in its size,
it might eventually grow into a large project over time. It is
therefore desirable to keep its code looks and feels consistent across
the different components, so that each member can communicate
smoothly.  Many software projects adopt ``style guides'', a set of
coding standards to maintain a certain level of superficial
consistency. While these guidelines help the project to achieve a
certain aspect of the code consistency, they mostly address the
stylistic aspects such as indentation or word capitalization. The
other types of consistency, or ``semantic consistency'' such as word
choice for concepts used in the code, is often left to each
developer's discretion.

One of the reasons why such a consistency is difficult to achieve is
that they are highly subjective. It is relatively easy to prescribe
the stylistic aspect of source code in a way that it can be checked
automatically. On the other hand, it is hard to regulate all the names
used in source code in advance. Programmers often encounter situations
where they have to name a certain concept that is not so well-defined,
yet necessary to be named. Some of these concepts might not be
immediately comparable to any real-life objects, so the programmers
have to be inventive. In general, naming is much more difficult to
regulate than superficial coding styles.

One way to cope with such a problem is to maintain a list of words
that are used for names and share them among the developers
\footnote{News organizations took a similar approach to this by
developing their in-house style guides \cite{ap_04, nyt_99}. }.
However, this is impractical because, unlike style guides, the
concepts used in a program are often project specific, and the same
word can mean different things in different projects.  For example,
the word ``{\tt view}'' can mean a portion of a table when it is used
in a database engine, while it can mean a visible window when it is
used in a graphical application. Or ``{\tt rate}'' can mean different
things in a financial application and a network application, etc. To
make it worse, different projects use different abbreviation for the
same words (e.g. ``{\tt att}'' versus ``{\tt attr}'' for
attributes). To date, there are still relatively few naming guidelines
for a large project.

Yet, the importance of names in a program code has been emphasized by
many researchers and practitioners \cite{kernighan_99, mcconnel_04}.
Programmers tend to heavily rely on meaningful identifier names to
understand source code \cite{lawrie_06b}, and they generally prefer a
long descriptive name than single-letter variables
\cite{beniamini_17}. It is also reported that poor naming can lead to
misunderstanding or confusion among programmers, which eventually
result in poor code quality \cite{avidan_17}. In many software
projects, inconsistent naming is often considered as a bad
smell. Sometimes they are actually treated as bugs ({\it naming bugs}
\cite{host_09}). While naming inconsistency does not immediately lead
to a malfunction, it decays the code quality over time and developers
become more prone to introduce serious bugs. A quick search over GitHub
reveals that programmers keep being confused by wrongly named
variables and methods.
\footnote{A search over GitHub issues with ``{\tt wrong name}'' turns
  out over 1 million results.}.
Also, it is natural to assume that the
problem of inconsistent names is exacerbated as the size of
codebase grows. In large software projects, multiple programmers are
involved in changing different parts of the code, sometimes with not
enough communication to each other. Without the holistic view of the
entire codebase, the inconsistent names can be often overlooked or
neglected for a long time, which further degrade the code quality.


Naming issues have been an active topic of research in the software
engineering community. Overall, two approaches exist; one is to
provide accurate names based on source code, and the other is to
detect and correct the naming inconsistency in existing code.  For
the first approach, Allamanis et al.  used machine learning algorithms
to suggest method and class names from code \cite{allamanis_15}. Alon
et al. converted source code into word embeddings \cite{mikolov_13}
that correspond to a certain word in natural language \cite{alon_19},
which can be used for identifiers. Raychev et al. recovered variable
names from obfuscated JavaScript code \cite{raychev_15}. For the
second approach, H\o{}st et al. \cite{host_09} used manually crafted
rules to detect naming bugs in a number of open source Java
projects. Liu et al. used machine learning to capture the
relationships between a method name and its body (code) to discover
bad method names that do not properly describe its function
\cite{liu_19}. Allamanis et al. proposed a method to automatically
capture the stylistic conventions from source code, some of them are
naming-related \cite{allamanis_14}. To our knowldege, our work is
the first attempt to detect semantic inconsistency of variable names
in a large software project.

As for the naming guidelines and code readability, \cite{basili_96}
offers an early work for the code documentation. Lawrie et
al. \cite{lawrie_06a} was one of the early attempts to give an insight
to naming bugs.  According to Lawrie et al., naming bugs are divided
into two categories: homonym and synonym. If two different concepts
are mapped into the same name (homonym), or a single concept is called
by multiple names (synonym), it often causes confusion to the
programmers \cite{deissenboeck_06}. Apart from naming, detecting
semantic inconsistency in source code can be a powerful tool for
finding potential bugs \cite{engler_01}.


\subsection{Importance of Variable Names}

So far, most of the existing works focus on the method names. This is
understandable because a method or function is often a meaningful
chunk of code that is supposed to have a coherent name. However, we
argue that it is a variable name, rather than a method name, that
plays the crucial role for understanding the high-level meaning of the
code because it reveals the type of information that the program
handles.  We have found that nouns often play a significant role in
naming variables, and in turn help the overall understanding of the
program. We also have found that each project has a fairly specific
set of nouns that are related to its target domain.

Informally, this can be shown in the two steps: First, method names
often rely on variable names. Table \ref{tab:related} shows the number
of method names that are related to the variables it uses. By
``related'', we mean that the method name contains a word that is also
used by one of the variables it uses. From this table, we can say that
about one third (or more) of method names rely on the variable names
in its meaning, i.e. we first need to understand the meaning of each
variable in order to understand the meaning of a method.

\begin{table}[tb]
  \caption{Number of Method Names}
  \vspace*{-1em}
  \label{tab:related}
  \centering
  \begin{tabular}{|l|r|r|} \hline
    Project & Related & Unrelated \\ \hline
    ant & 5,507 & 5,905 \\
    antlr4 & 1,046 & 1,922 \\
    bcel & 1,113 & 2,243 \\
    compress & 995 & 1,366 \\
    jedit & 3,068 & 4,630 \\
    jhotdraw & 2,171 & 5,127 \\
    junit4 & 396 & 943 \\
    lucene & 3,978 & 8,026 \\
    tomcat & 10,519 & 11,189 \\
    weka & 11,040 & 11,428 \\
    xerces & 3,820 & 4,272 \\
    xz & 224 & 464 \\
    \hline
  \end{tabular}
\end{table}

Next, many variable names are made up with nouns that refer to domain
specific concepts. Note that many method names consist of both verbs
{\it and} nouns.  Again, this is understandable as a method is often
an actor or agent of the objects while a variable usually contains a
reference to certain objects. However, we observed that the verbs used
in method names are fairly limited in its variety. Table
\ref{tab:topwords} shows the popular words used in method names. One
can see that the most popular verbs are {\tt get}, {\tt set}, {\tt
  add}, {\tt remove}, and {\tt create} in nearly all projects, whereas
the nouns are more diverse across different projects. One can also see
that the nouns are often specific to each project topic, whereas the
verbs are mostly generic.

\begin{table}[tb]
  \caption{Popular Words Used in Method Names}
  \vspace*{-1em}
  \label{tab:topwords}
  \centering
  \setlength \tabcolsep{4 pt}
  \begin{tabular}{|l|p{10em}|p{10em}|} \hline
    Project & Top Verbs & Top Nouns \\ \hline
    ant & {\scriptsize
      {\tt set}, {\tt get}, {\tt add}, {\tt create}, {\tt is}
    } & {\scriptsize
      {\tt file}, {\tt name}, {\tt function}, {\tt class}, {\tt output}
    } \\ \hline
    antlr & {\scriptsize
      {\tt get}, {\tt set}, {\tt add}, {\tt remove}, {\tt visit}
    } & {\scriptsize
      {\tt string}, {\tt rule}, {\tt token}, {\tt code}, {\tt name}
    } \\ \hline
    bcel & {\scriptsize
      {\tt visit}, {\tt get}, {\tt accept}, {\tt set}, {\tt dump}
    } & {\scriptsize
      {\tt constant}, {\tt class}, {\tt string}, {\tt type}, {\tt value}
    } \\ \hline
    compress & {\scriptsize
      {\tt get}, {\tt set}, {\tt read}, {\tt write}, {\tt close}
    } & {\scriptsize
      {\tt stream}, {\tt entry}, {\tt archive}, {\tt data}, {\tt input}
    } \\ \hline
    jedit & {\scriptsize
      {\tt get}, {\tt set}, {\tt add}, {\tt is}, {\tt run}
    } & {\scriptsize
      {\tt jj}, {\tt action}, {\tt line}, {\tt string}, {\tt buffer}
    } \\ \hline
    jhotdraw & {\scriptsize
      {\tt get}, {\tt set}, {\tt create}, {\tt is}, {\tt add}
    } & {\scriptsize
      {\tt action}, {\tt figure}, {\tt color}, {\tt name}, {\tt property}
    } \\ \hline
    junit4 & {\scriptsize
      {\tt get}, {\tt assert}, {\tt run}, {\tt test}, {\tt validate}
    } & {\scriptsize
      {\tt test}, {\tt class}, {\tt method}, {\tt failure}, {\tt runner}
    } \\ \hline
    lucene & {\scriptsize
      {\tt get}, {\tt set}, {\tt compare}, {\tt add}, {\tt read}
    } & {\scriptsize
      {\tt doc}, {\tt next}, {\tt string}, {\tt value}, {\tt bytes}
    } \\ \hline
    tomcat & {\scriptsize
      {\tt get}, {\tt set}, {\tt is}, {\tt add}, {\tt remove}
    } & {\scriptsize
      {\tt name}, {\tt string}, {\tt session}, {\tt max}, {\tt class}
    } \\ \hline
    weka & {\scriptsize
      {\tt get}, {\tt set}, {\tt add}, {\tt is}, {\tt create}
    } & {\scriptsize
      {\tt text}, {\tt tip}, {\tt options}, {\tt action}, {\tt string}
    } \\ \hline
    xerces & {\scriptsize
      {\tt get}, {\tt set}, {\tt is}, {\tt create}, {\tt add}
    } & {\scriptsize
      {\tt element}, {\tt name}, {\tt decl}, {\tt type}, {\tt impl}
    } \\ \hline
    xz & {\scriptsize
      {\tt get}, {\tt write}, {\tt read}, {\tt close}, {\tt set}
    } & {\scriptsize
      {\tt stream}, {\tt input}, {\tt size}, {\tt output}, {\tt memory}
    } \\ \hline
  \end{tabular}
\end{table}

From the above observations, we can conclude the following:

\begin{enumerate}
\item Nouns often play a significant role in naming identifiers,
  and in turn understanding the high-level meaning of the code.
\item Each project has a fairly specific set of nouns
  that are related to its target domain.
\end{enumerate}

Relatively fewer attempts have been made for predicting field or
variable names, possibly because of its variety and subjectivity.
Typically, nouns are used for variable names whereas verbs are used for
method names. WordNet \cite{miller_95} has 115k nouns while it has
only 11k verbs.  The software industry has a long history of using
existing common nouns for representing abstract concepts, such as
``tree'', ``view'' or ``stream''. The meaning of these words, however,
are only loosely defined in each project. Raychev et al. obtained the
characteristics of variable names from a large JavaScript codebase
\cite{raychev_15}, which can be applied to general functions but not
project specific ones.

There is another reason why we think variable names are important:
ultimately, what a computer program handles is just a collection of
bits; they are typically interpreted as numbers, vectors and strings.
However, that is not the end -- real applications need to handle
concepts such as ``counter'', ``position'' or
``balance''.
Fig. \ref{fig:twofuncs} shows two functions that are
functionally identical (adding a value of one variable to another) but
semantically different; one is to update the balance, and the other is
to update the position.
Similarly, a string can be used as ``username'', ``pathname'' or ``address''.
In other words, they need
to assign {\it the meaning} to these bits by naming them, and it is
the primary function of variables.  Variable names are particularly
important to give programmers high-level views.  They represent a
fundamental building block of application domain.

\begin{figure}[tb]
\centering
\begin{tabular}{|lp{20em}|} \hline
{\bf a.} &
\begin{minipage}[t]{\linewidth} \scriptsize
\begin{verbatim}
void update(float income) {
  balance += income;
}
\end{verbatim}
\end{minipage}
\\ \hline
{\bf b.} &
\begin{minipage}[t]{\linewidth} \scriptsize
\begin{verbatim}
void update(float velocity) {
  position += velocity;
}
\end{verbatim}
\end{minipage}
\\ \hline
\end{tabular}
\caption{Functionally Identical But Semantically Different Functions}
\label{fig:twofuncs}
\end{figure}

In reality, maintaining consistency is not always a project's top
goal.  Real world software faces constant challenge to be modified or
improved.  Sometimes one needs to break consistency in order to
upgrade a part of the code to adapt for a newer requirement.
Therefore, the naming rules are often a set of conventions rather than
a strict dogma. Our goal is to capture these conventions and reuse
them effectively in an automated manner to improve the code quality.
In this paper, we first present our general framework of testing
consistency, and then introduce an actual mechanism to apply it
to source code.


\section{What is Consistency?}
\label{sec:consistency}



In this section, we present a general framework of testing the
consistency between two sets of inputs over a certain invariant.
Suppose we have two sentences, $S_a$ and $S_b$, where each sentence
has a pair of features $(F_{a1}, F_{a2})$ and $(F_{b1}, F_{b2})$,
respectively.  Furthermore, assume that $F_{a1}$ in general conveys
a sufficient context to reliably predict $F_{a2}$, i.e. there
is a function $K$ such that $K(F_{a1}) = F_{a2}$.  Now, if we also
find that $K(F_{b1}) = F_{b2}$, i.e. $F_{b1}$ has the same context to
predict $F_{b2}$, we can say that $S_b$ is consistent with $S_a$ in
regard to $K$.

To illustrate this framework, take a look at the following example:

\begin{enumerate}[label=(\alph*)] 
  \item ``{\tt It is \underline{rainy} today so you should take an \underline{umbrella}.}''
  \item ``{\tt It is \underline{[X]} today so John should take an \underline{umbrella}.}''
\end{enumerate}

Suppose that we obtained the knowledge ($K$) that there is a strong
relationship between ``{\tt rainy}'' and ``{\tt umbrella}'' from
sentence (a).  It is fairly easy then to predict the word {\tt [X]}
according to our knowledge.  If the word {\tt [X]} is indeed ``{\tt
  rainy}'', sentence (b) is consistent with sentence (a) in regard to our
knowledge, $K$. While this formulation is similar to a typical machine
learning framework, the focus is different: instead of predicting the
unknown value of {\tt [X]}, we are interested in measuring the
consistency of $K$ over various inputs.

In reality, however, this kind of categorical knowledge is hard to
obtain. Therefore, we extend our definition to include Bayesian
inference, i.e. statements that have varying degrees of
certainty. Suppose we have two statements, $S_a$ and $S_b$, pairs of
their features $(F_{a1}, F_{a2})$ and $(F_{b1}, F_{b2})$.  Now, if we
find that $F_{a2}$ is {\em likely to be predicted by} $F_{a1}$,
i.e. $P(F_{a2} | F_{a1})$ is high, and we also find that $P(F_{b2} |
F_{b1})$ is high, we can say that $S_b$ is likely to be consistent
with $S_a$ in regard to $P$. In other words, the consistency in our
framework is equivalent to the {\em predictability} of answers.

\subsection{Measuring Consistency of Program}

Now, let us apply the above framework to a program code. Suppose we
have two comparable code snippets A and B
(Fig. \ref{fig:snippets}). In snippet A, the name ``{\tt out}'' is used
for a return value of {\tt open(...)}  function and also for the first
argument of {\tt write(...)} function.  Assume that we learned the
relationship between the variable name ``{\tt out}'' and these two
statements. More formally put, we find that the snippet A has
three features:

\begin{itemize} 
  \item $F_{A1}$: the variable is assigned with the return value of ``{\tt open()}''.
  \item $F_{A2}$: the variable is passed as the first argument of ``{\tt write()}''.
  \item $F_{A3}$: the variable has name ``{\tt out}''.
\end{itemize}

We might say that this is the ``knowledge'' $K$ that we learned about
the use of this variable. Furthermore, we have another code snippet B
where a certain variable {\tt X} is used in the exactly same manner;
it has three features and we find $F_{A1}$ = $F_{B1}$ and $F_{A2}$ =
$F_{B2}$. If we find $F_{A3}$ is also equal to $F_{B3}$, i.e. the name
of the variable {\tt X} is indeed ``{\tt out}'', we can say that the
variable name is consistent in regard to our knowledge $K$. The key
idea here is that most variables exist with relationship with other
variables, and the relationship defines the role (name) of each
variable \footnote{ We assume that variables with a similar role tend
  to have a similar name. } which can be inferred from various aspects
of source code.

\begin{figure}[tb]
\centering
\setlength \tabcolsep{8 pt}
\begin{tabular}{p{10em}|p{10em}}
\begin{minipage}[b]{\linewidth} \scriptsize
{\bf Snippet A:} \\ \tt
\underline{out}\verb+ = open(...);+ \\
\verb+write(+\underline{out}\verb+, ...);+ \\
\end{minipage}
&
\begin{minipage}[b]{\linewidth} \scriptsize
{\bf Snippet B:} \\ \tt
\underline{X}\verb+ = open(...);+ \\
\verb+write(+\underline{X}\verb+, ...);+ \\
\end{minipage}
\\
\end{tabular}
\caption{Comparable Code Snippets}
\label{fig:snippets}
\end{figure}

So far, the framework we presented here is general in that we did not
put any assumption on how each feature should look like or what their
relationship can be.  We later create a more concrete mechanism to
express a usage of a variable, and a probabilistic model (knowledge)
to measure its semantic consistency in a similar process described
above. If the program is not consistent with our model, we can suggest
a better name for variables that aligns with our understanding of the
program. However, the general framework can be applied to any kind of
elements in source code. For example, it is possible to measure the
consistency between method names and its calling convention, if such
features are available. In the rest of this paper, however, we focus
on improving the consistency of variable names to improve the code
readability using the above framework.


\section{Proposed Method}
\label{sec:useflowgraph}

\begin{figure}[tb]
\centering

\setlength \tabcolsep{4 pt}
\begin{tabular}{|p{6em} p{6em}|} \hline
\vspace*{-10em}
\begin{minipage}[b]{\linewidth} \scriptsize
\begin{verbatim}
f(a, b, c)
{
  x = (a+b) * c;
  y = -a;
  return x + y;
}
\end{verbatim}
\end{minipage}
&
\begin{minipage}[b]{\linewidth}
\includegraphics[width=\linewidth]{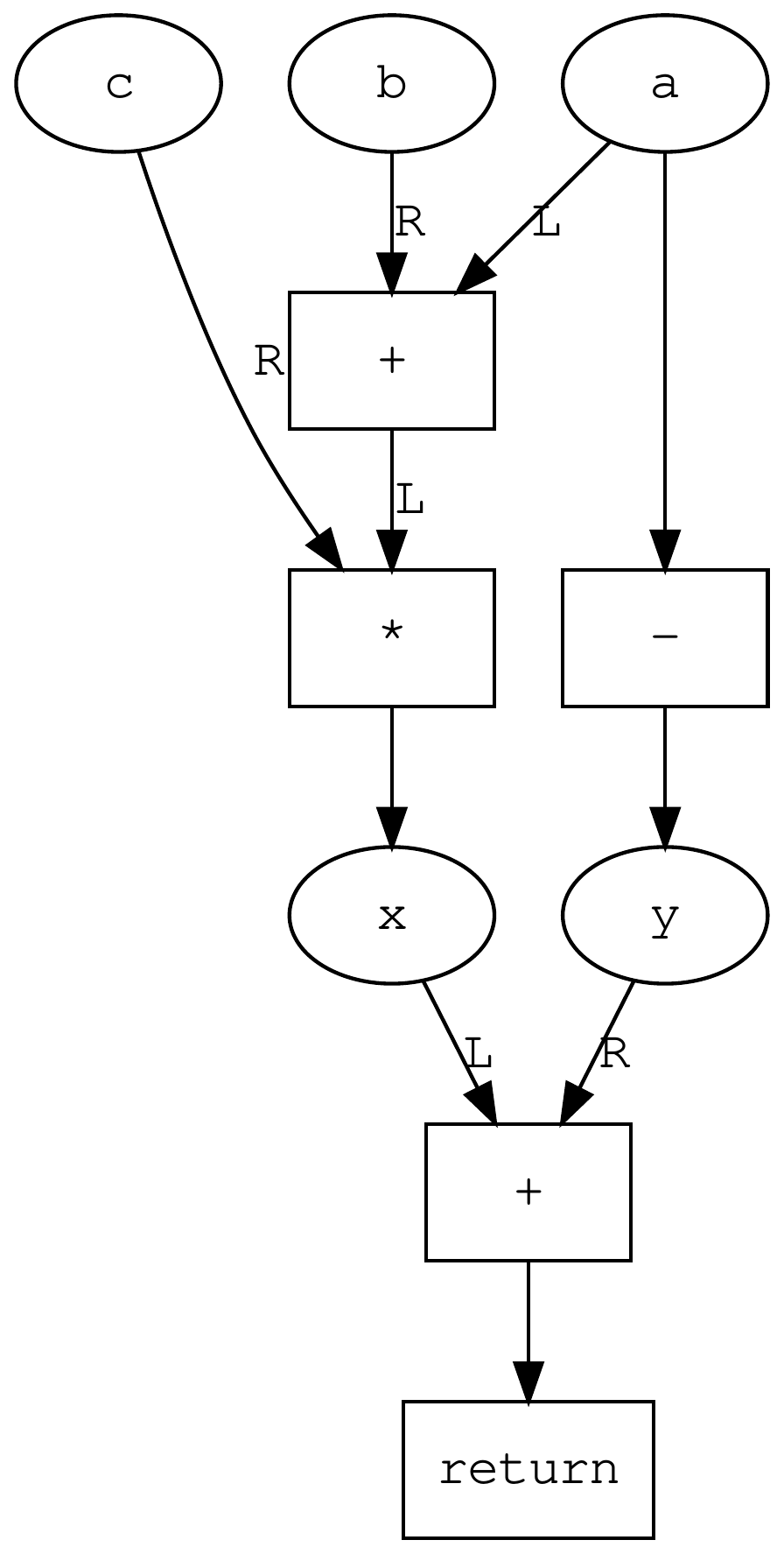}
\end{minipage}
\\ \hline
\end{tabular}

\caption{Simple Use-Flow Graph (UFG)}
\label{fig:ufg_basic}
\end{figure}
\begin{figure}[tb]
\centering
\fbox{\includegraphics[width=8em]{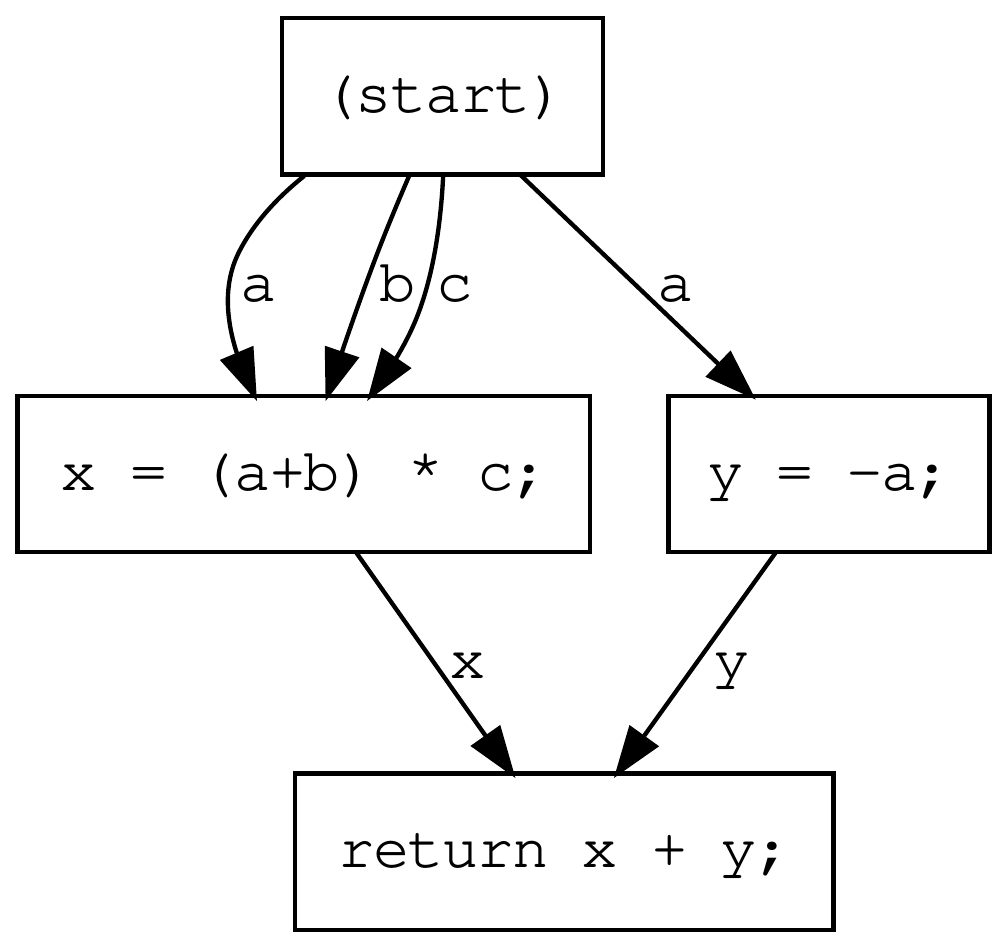}}
\caption{Typical Program Dependence Graph (PDG)}
\label{fig:pdg_basic}
\end{figure}

In this section, we describe how to apply the above framework of
naming consistency to program variables. First, we need to capture the
usage of each variable in a systematic way. For this purpose, we
introduce a graph structure called ``Use-Flow Graph'' (UFG).  The idea
of UFG is similar to a dataflow diagram and program dependence graph
(PDG). A typical dataflow diagram describes how data is processed and
transmitted from one part of a system to another.  In most settings,
the parts involved in a dataflow diagram are processors or storage
devices. UFG involves with a more granular kind of storage: variables
and fields in a program. In this sense, UFG is similar to a
PDG. However, while a typical PDG only shows the data dependence of
each {\it statement}, UFG shows the data dependence between each {\it
  variable}.  The idea of using a graph for representing the dataflow
among variables was disseminated by \cite{reps_97}.  We added
operators and function (method) arguments as a location.
Fig. \ref{fig:ufg_basic} shows a sample UFG. Note that the graph not only shows
how the value is transmitted from each variable ({\tt a}, {\tt b},
{\tt c}, {\tt x} and {\tt y}), but also shows how various operators ({\tt +},
{\tt -} and {\tt *}) are applied in the process.  This way, we can see
how the value of each variable is treated in a series of processing
\footnote{
  One of the major differences between our graph and the work by Raychev et al.
  \cite{raychev_15} is that our graph is
 directional; we only consider a relationship that reflects an actual
 execution order. For example, there is no direct relationship between
 variable {\tt x} and {\tt y} in this graph.
}.
A comparable PDG for the same program
could be written as in Fig. \ref{fig:pdg_basic}.

We further added a way to express conditional branches and loops to
UFG, which is explained later.  In short, UFG can present how values
(variables, fields or constants) are treated at each operation in a
precise manner without depending on a language syntax.  A path in UFG
can show how a particular value is given as an input, processed and
tested, and passed to other variables. We then define the ``usage
pattern'' of a variable as a UFG path that is originating from that
variable.  By traversing the edges in Fig. \ref{fig:ufg_basic},
we obtain the following usage patterns:

\begin{itemize} 
\item \verb+a+ $\rightarrow^L$ \verb.+. $\rightarrow^L$ \verb.*. $\rightarrow$ \verb.x. $\rightarrow^R$ \verb.+. $\rightarrow$ \verb+return+
\item \verb+a+ $\rightarrow$ \verb.-. $\rightarrow$ \verb.y. $\rightarrow^L$ \verb.+. $\rightarrow$ \verb+return+
\item \verb+b+ $\rightarrow^R$ \verb.+. $\rightarrow^L$ \verb.*. $\rightarrow$ \verb.x. $\rightarrow^R$ \verb.+. $\rightarrow$ \verb+return+
\item \verb+c+ $\rightarrow^R$ \verb.*. $\rightarrow$ \verb.x. $\rightarrow^R$ \verb.+. $\rightarrow$ \verb+return+
\end{itemize}

After obtaining such patterns, we construct and use a probabilistic
model that we explained in Section \ref{sec:consistency} to test if
each variable name is consistent with its usage pattern. In the next
subsection, we first explain how to construct UFGs from source code.

\subsection{Constructing Use-Flow Graph}

We now illustrate how to construct a UFG from typical language constructs in Java
\footnote{
  Our current UFG generator fully supports Java 8 syntax.
  In future, we plan to extend this to other popular procedural languages such as
  C\# or C++.
}. UFG can be constructed in a linear time for a given program size,
allowing to analyze a large project in a reasonable time.

Let us revisit the UFG shown in Fig. \ref{fig:ufg_basic}.  In this
graph, every operator is represented as a separate node, and the
transmission of each value is shown as directed edges. The label of
each edge shows at which side of the binary operator that a value is
used (either {\tt L} or {\tt R}).  Note that the order of execution
can be recovered by following the edges at each node and each variable
is still distinguished as a different path in a graph. So it is still
possible to reconstruct the equivalent program from a given graph
\footnote{Exact reconstruction of the original code is not always
  guaranteed, because not all the side effects and indirect access are
  preserved. We assume that the lack of these properties do not cause
  a significant loss of accuracy for our purposes in this paper.}.  We
expect that the overall structure of UFG is generally preserved across
different programming styles
\footnote{
  Note that we do not intend to identify the functional equivalence.
  Two mathematically equivalent
  expressions (e.g. {\tt a+b} and {\tt b+a}) does not necessarily
  result in the same UFG.  Our goal here is to preserve the
  intent of programmers as much as possible while removing stylistic
  differences.
}, because
all the basic operations still have to be applied
in the same order to have the same effect.
The UFG of a program shows how each piece of data at various
locations in a program is interacted with each other.
If one takes a look at the UFG around a certain variable,
its subgraph is likely to show how the variable is used at the other parts of the program;
namely, they are showing its {\it usage}.

\subsection{Tracking Multiple Variables}

\begin{figure}[tb]
\centering
\setlength \tabcolsep{4 pt}
\begin{tabular}{|p{6em} p{7em}|} \hline
\vspace*{-6em}
\begin{minipage}[b]{\linewidth} \scriptsize
\begin{verbatim}
{
  x = x + y;
  z = y;
  a = 42;
}
\end{verbatim}
\end{minipage}
&
\begin{minipage}[b]{\linewidth}
\includegraphics[width=\linewidth]{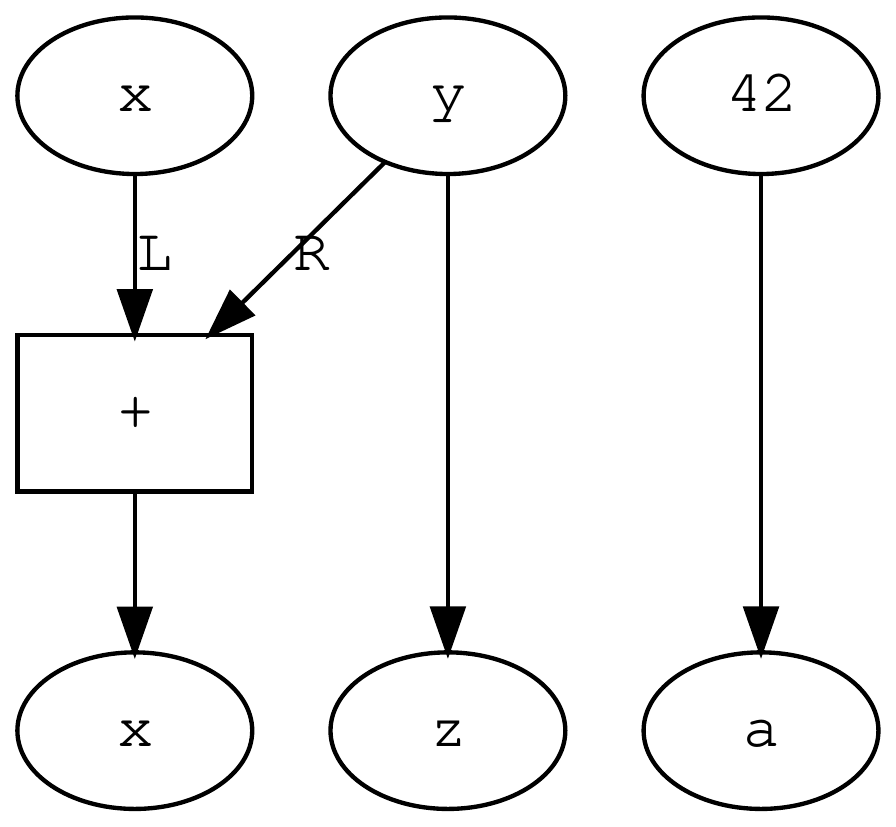}
\end{minipage}
\\ \hline
\end{tabular}
\caption{Use-Flow Graph of Multiple Variables}
\label{fig:ufg_side}
\end{figure}

When a program is purely functional, i.e. its output is solely
determined by its inputs and there is no side effect, the program can
be represented by a single connected UFG with one sink
node. When multiple independent variables are modified, however, there
will be multiple sinks or disjointed graphs, as shown in Fig. \ref{fig:ufg_side}.
A different graph concerns a different set
of data that are unrelated to each other. Note that the original order
of execution is not preserved because there is no dependence between
statements, and these unrelated statements could be executed
in parallel.

\subsection{Conditional Statement}

\begin{figure}
\centering

\begin{multicols}{2}
\setlength \tabcolsep{4 pt}
\begin{tabular}{|p{3em} p{7em}|} \hline
\vspace*{-6em}
\begin{minipage}[b]{\linewidth} \scriptsize
\begin{verbatim}
if (x) {
  y = 1;
} else {
  y = 2;
}
\end{verbatim}
\end{minipage}
&
\begin{minipage}[b]{\linewidth}
\includegraphics[width=\linewidth]{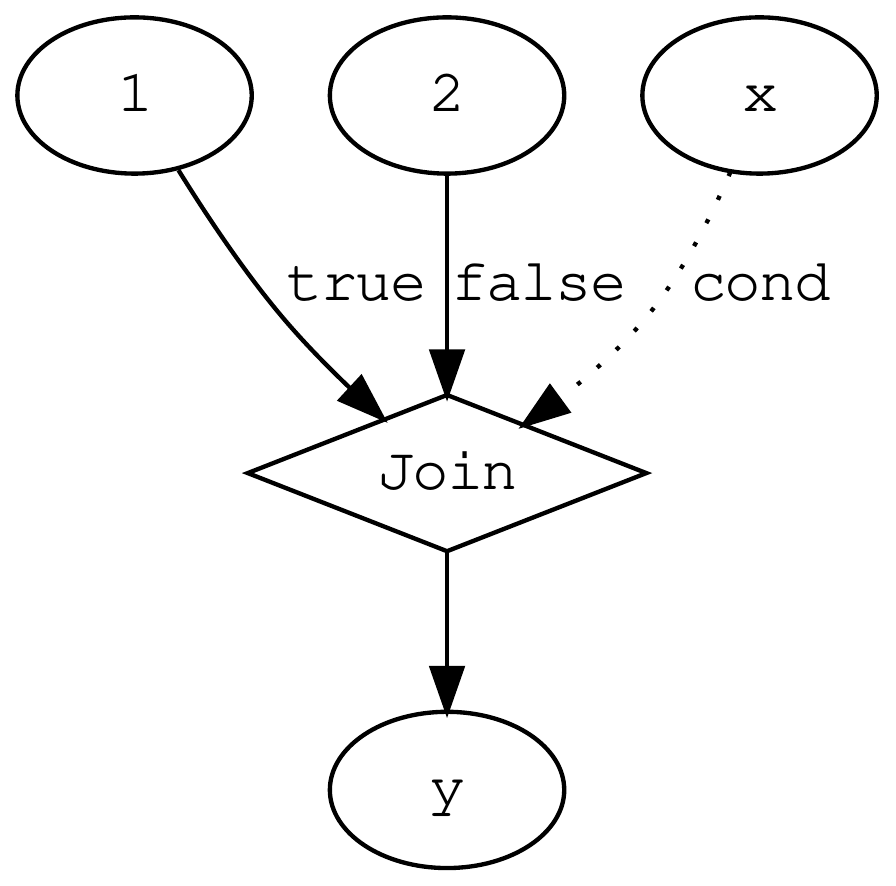}
\end{minipage}
\\ \hline
\end{tabular}
\caption{Use-Flow Graph of Conditional Statement}
\label{fig:ufg_cond}
\par

\setlength \tabcolsep{4 pt}
\begin{tabular}{|p{3em} p{6em}|} \hline
\vspace*{-6em}
\begin{minipage}[b]{\linewidth} \scriptsize
\begin{verbatim}
x = f(
 y,2,3
)
\end{verbatim}
\end{minipage}
&
\begin{minipage}[b]{\linewidth}
\includegraphics[width=\linewidth]{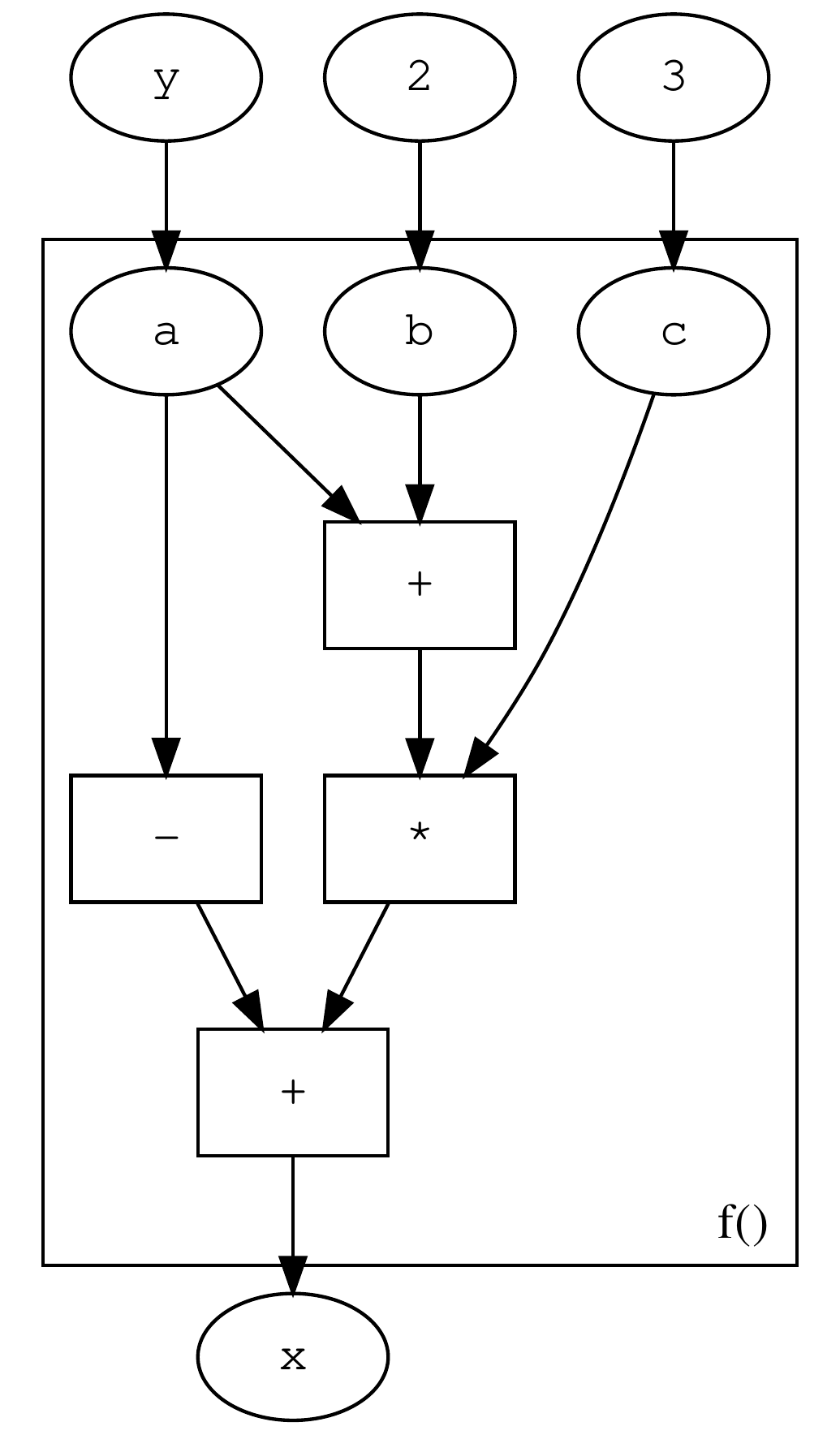}
\end{minipage}
\\ \hline
\end{tabular}
\caption{Use-Flow Graph of Embedded Function Call.
  (Edge labels are removed for readability.)}
\label{fig:ufg_embed}
\end{multicols}
\end{figure}

To represent conditional statements such as {\tt if}, we introduce
special nodes. When a value of a certain variable is determined
conditionally, all the possible flows are connected to a single {\tt
  Join} node (Fig. \ref{fig:ufg_cond}). The idea is to interpret a
{\tt Join} node as something like a railroad switch, or a conditional
operator.  When the conditional statement is executed, only one of
these edges ({\tt true} or {\tt false} in this example) is used. Each
edge is labeled with its condition so that they can still be
distinguished.  When multiple variables are modified in the
{\tt if} statement, a similar structure is created for every variable
that changed. Note that each statement is converted to nodes in UFG
whether or not the statement is actually executed.

\subsection{Loop}

\begin{figure}[tb]
\centering
\setlength \tabcolsep{4 pt}
\begin{tabular}{|p{7em} p{7em}|} \hline
\vspace*{-10em}
\begin{minipage}[b]{\linewidth} \scriptsize
\begin{verbatim}
do {
  S;  // modify x
} while (p);
\end{verbatim}
\end{minipage}
&
\begin{minipage}[b]{\linewidth}
\includegraphics[width=\linewidth]{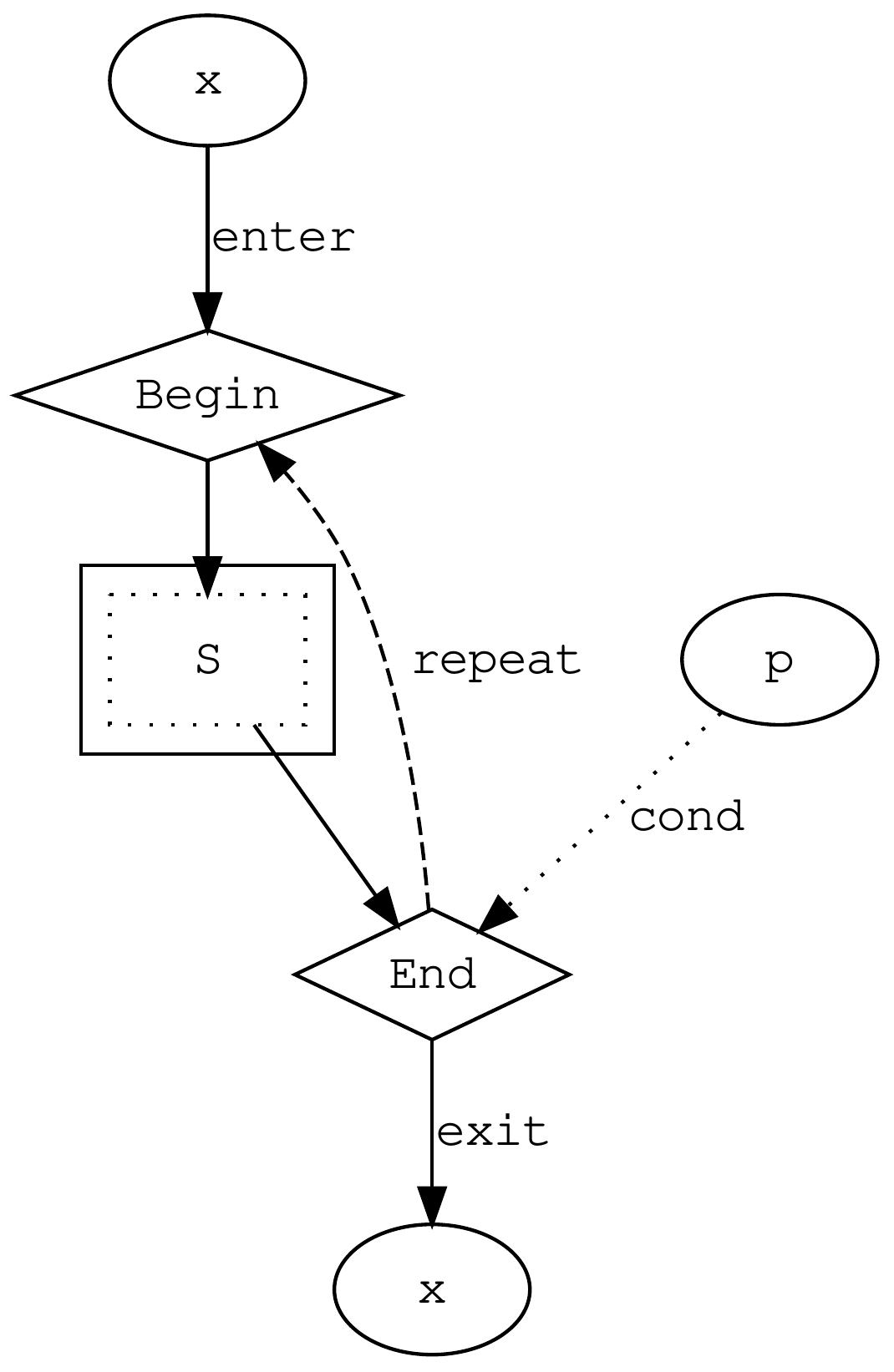}
\end{minipage}
\\ \hline
\end{tabular}
\caption{Use-Flow Graph of Loop.
  {\tt S} denotes an inner subgraph that modifies variable {\tt x}.
}
\label{fig:ufg_loop}
\end{figure}

We introduce another set of special nodes, {\tt Begin} and {\tt End},
to represent a loop. (Fig. \ref{fig:ufg_loop}).  For each variable
that is modified in the loop, its Use-Flow subgraph is sandwiched with
{\tt Begin} and {\tt End} nodes. In the case of {\tt do} loop, as
shown in Fig. \ref{fig:ufg_loop}, the conditional test is performed at
the end of the loop, and the {\tt End} node behaves like the {\tt
  Join} node in the previous example. The interpretation of this graph
is that the subgraph between the {\tt Begin} and {\tt End} nodes are
repeated by an unknown number of times, and for every time the
conditional value {\tt p} is reevaluated. Note that the purpose of
this graph is to show in {\it which context} each variable is
modified, but not to show how the loop actually runs. A UFG is not
suitable for inferring the loop invariant or comparing different loop
structures.

\subsection{Function/Method Call}

\begin{figure}[tb]
\centering
\setlength \tabcolsep{4 pt}
\begin{tabular}{|p{4em} p{7em}|} \hline
\vspace*{-5em}
\begin{minipage}[b]{\linewidth} \scriptsize
\begin{verbatim}
x = f(y,2,3)
\end{verbatim}
\end{minipage}
&
\begin{minipage}[b]{\linewidth}
\includegraphics[width=\linewidth]{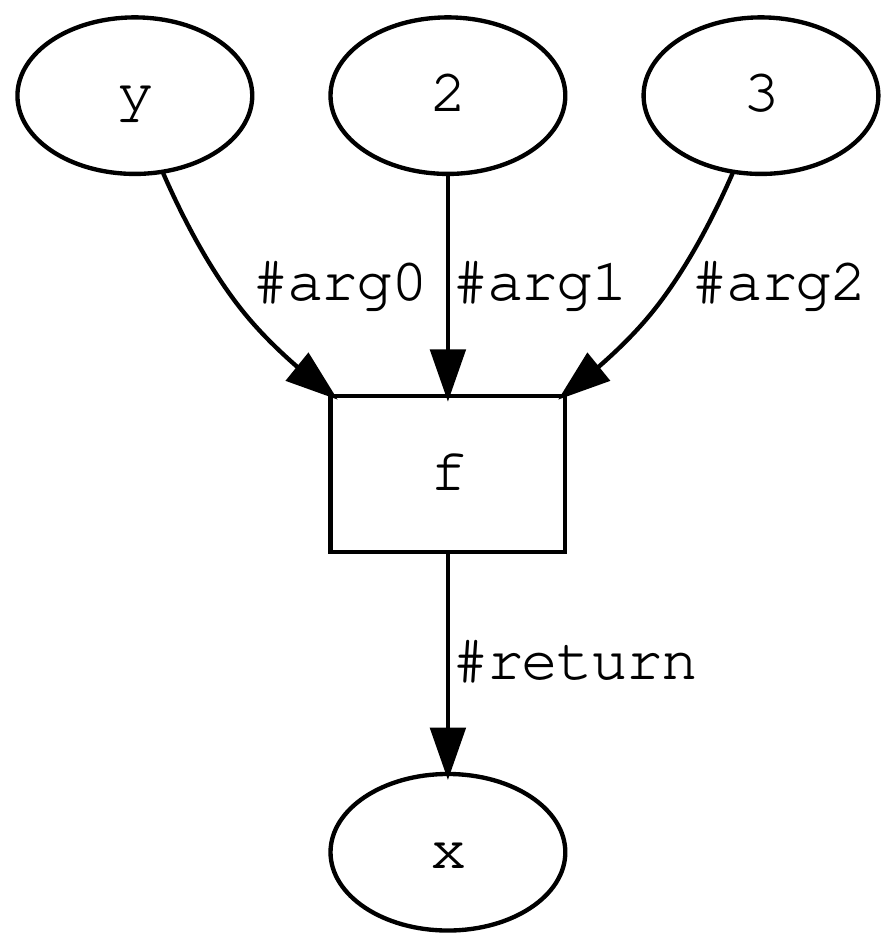}
\end{minipage}
\\ \hline
\end{tabular}
\caption{Use-Flow Graph of Referenced Function Call}
\label{fig:ufg_funcall}
\end{figure}

A function or method call in a UFG is simply treated as yet another
operator node (Fig. \ref{fig:ufg_funcall}).
The callee function is {\it referenced}.
Each function call node has all possible references
to the functions that has the same signature, including virtual
functions.  Each edge for the function arguments is labeled as {\tt
  \#arg0}, {\tt \#arg1} and {\tt \#arg2} and the return value as {\tt
  \#return}.  When we want to obtain a relationship of nodes across
multiple functions, however, we can internally treat each call node as
if there was another UFG embedded within the node, in a similar manner
to code inlining, i.e. the callee function is {\it embedded}
(Fig. \ref{fig:ufg_embed}).  This allows us to consider the
relationship of value operations in an interprocedural context.

\subsection{Collecting Usage Patterns}

Realistic software usually contains thousands of functions.  Since we
want to collect a usage pattern of a value in a long context, we want
to track how the value is handled and passed across multiple
functions.  This is done in the following steps. The key idea here is
to start from a node of interest (a variable in this case) and
gradually incorporate other nodes to multiple functions that are being
called:

\begin{enumerate} 
\item Start from every variable node (a node referring to a variable).
  This is an initial pattern for this variable usage \footnote{
    The initial variable node itself is not included in a usage pattern.}.
\item Pick the next node by tracing its outgoing edges.
  Incorporate it as a part of the pattern.
\item If the node is a function call, push the current node to the stack
  and connect a value node of the caller function to the corresponding argument node
  of the callee function \footnote{
    In this paper, we limit the number of possible virtual functions
    that are potentially referenced to 5. When there are 6 or more possible
    virtual functions, we picked the virtual methods for the five most specific class. }.
  This is done by connecting the UFGs of both functions.
  Incorporate this connection to the pattern.
\item If the node is a return node and the stack {\it is not} empty,
  pop the previous node from the stack.  Connect the return node of
  the callee function with the receiving node of the caller function.
\item If the node is a return node and the stack {\it is} empty,
  Connect the return node of the callee function with the receiving node
  of a function call node for {\it every} function that it potentially
  calls that function. Multiple patterns are generated.
\item Repeat this process until a pattern grows to a maximum predefined length \footnote{
  Note that an exponential number of patterns can be generated
  when there is a branch in the graph.
  We limited the number of maximum nodes to be traced to five. }
\end{enumerate}

There are actually two kinds of usage patterns: forward and backward.
The process described above is one for obtaining forward usage
patterns.  For backward patterns, the same process is used for the
opposite direction of the edges. From now on, we just use the term
``usage patterns'' for forward and backward patterns, combined.

Let us illustrate the above algorithm with a more realistic example
(Fig. \ref{fig:sample_java}).  Suppose that we are interested in
taking a usage pattern of the variable ``{\tt line}'' at Line 4. We
start from the assignment expression, incorporate the {\tt .indexOf()}
and {\tt .substring()} node, and reach the end of the {\tt getName()}
function at the {\tt return} statement. Then we further extend the
pattern by incorporating the nodes that are receiving the value of
{\tt getName()} function. In this example, the {\tt name} node is added to
the pattern. We repeat the same process for the backward pattern.  At
the end, we obtain the following usage pattern for the variable ``{\tt
  line}''.

\begin{itemize} 
\item \verb+fp.readLine()+ $\rightarrow$ \underline{\tt line} $\rightarrow^\mathit{this}$ \verb.indexOf(). $\rightarrow^\mathit{arg1}$ \verb.substring(). $\rightarrow$ \verb.name. $\rightarrow^L$ \verb.+. $\rightarrow^\mathit{arg0}$ \verb.println().
\end{itemize}

\begin{figure}[tb]
\centering
\begin{minipage}[t]{0.8\linewidth}
\begin{lstlisting}[frame=single, numbers=left, basicstyle=\scriptsize\ttfamily, escapechar=@]
private BufferedReader fp;
public String getName() {
  String @{\underline{line}}@ = fp.readLine();
  int i = line.indexOf(' ');
  return line.substring(0, i);
}
public void show() {
  String name = @{\underline{getName()}}@;
  System.out.println(name+"!!");
}
public String getField() {
  String @{\underline{buf}}@ = fp.readLine();
  return buf.substring(0, buf.indexOf(':'));
}
public String getColumn() {
  String @{\underline{buf}}@ = fp.readLine();
  return buf.substring(0, buf.indexOf(','));
}
\end{lstlisting}
\end{minipage}
\caption{Sample Java Code}
\label{fig:sample_java}
\end{figure}

\subsection{Detecting Inconsistent Names}

In the snippet shown in Fig. \ref{fig:sample_java}, there are other
functions named {\tt getField()} and {\tt getColumn()}. We obtain
the usage pattern of their variables ``{\tt buf}'' as follows:

\begin{itemize} 
\item \verb+fp.readLine()+ $\rightarrow$ \underline{\tt buf} $\rightarrow^\mathit{this}$ \verb.indexOf(). $\rightarrow^\mathit{arg1}$ \verb.substring().
\end{itemize}

The above pattern is similar to the one obtained for the variable
``{\tt line}''. However, this pattern appears more frequently
throughout the program than the previous one, hence the pattern is
more strongly associated with the name ``{\tt buf}'' rather than
``{\tt line}''.  This way, the system can learn that ``{\tt line}''
should be better named as ``{\tt buf}'' to achieve more consistency.
Note that this sort of knowledge is acquired entirely from the project
source code. The system does not use any external knowledge.
Naturally, this allows the system to tune to a specific project.

To recapitulate, the overall algorithm of our proposed method is the
following:

\begin{enumerate} 
\item Extract UFGs from source code and
  collect the usage patterns for each variable.
\item Construct a probabilistic model to test
  if each variable name is consistent with its usage pattern.
\item If a variable name is found not to be consistent, suggest an
  alternative name that is more strongly associated with its usage
  pattern.
\end{enumerate}

\subsection{Constructing and Using Probabilistic Model}


After extracting UFGs from source code and obtaining its usage
patterns for each variable, we construct a probabilistic model.  We
try to learn a model that predicts a variable name from a given usage
pattern as explained in Section \ref{sec:consistency}. In this paper,
we used a simple Bayesian inference, i.e. we assume that every node in
a usage pattern independently affects the choice of its variable
name. In the case of the previous Java example, the features that
influence the prediction include: the origin of the value, the way it
is used, and its destination (the variable name it is assigned), and
so on. 
A usage pattern is converted to a
set of features by encoding the sequence of its adjacent node pairs.

For example, a usage pattern like this

\begin{itemize}
\item \verb.indexOf(). $\rightarrow^\mathit{arg1}$ \verb.substring(). $\rightarrow$ \verb.name. $\rightarrow^L$ \verb.+. $\rightarrow^\mathit{arg0}$ \verb.println().
\end{itemize}

is converted into the following features:

{
\begin{itemize}
\item \verb.indexOf():arg1:substring().
\item \verb.substring():name.
\item \verb.name:L:+.
\item \verb.+:arg0:println().
\end{itemize}
}

To give the model more flexibility, names are not treated as a single
feature but a set of features based on its tokens. For example, ``{\tt
  outputBufferName}'' is tokenized into three distinct features:
``{\tt output}'', ``{\tt buffer}'' and ``{\tt name}''. \footnote{ Our
  tokenizer assumes that variable names are either the form of {\tt
    camelCase} or {\tt snake\_case}.  }. Other than tokenization, the
system does not have any prior knowledge about the natural language
used in variable names.
The list of features for each node of a usage
pattern is shown in Table \ref{tab:features}.

After learning the model using all the usage patterns throughout the
program, the system re-applies them to every variable and see if its
prediction matches its original name. If it does, its name is
consistent with its usage. If it does not, the name is ``outvoted''
by other variables which have the same usage pattern. In this case,
the system suggests the most likely name for that usage pattern.

\begin{figure}[tb]
\centering
\fbox{\includegraphics[width=0.9\linewidth]{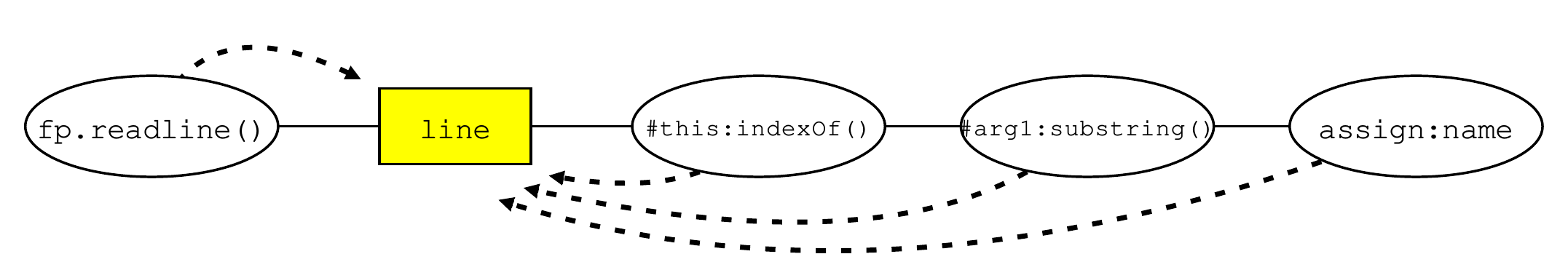}}
\caption{Probabilistic Model for Variable ``{\tt line}''}
\label{fig:prob_model}
\end{figure}

\begin{table}[tb]
  \caption{Features Included in Usage Patterns}
  \vspace*{-1em}
  \label{tab:features}
  \centering
  \fbox{\begin{minipage}[t]{0.8\linewidth}
\begin{enumerate}
\item If the value is a constant.
\item If the value is used within a loop.
\item If the value is used in a branch condition.
\item If the value is used as an array index.
\item If the value is used as an object instance.
\item Operator the value is used against.
\item Function argument the value is passed to.
\item Type of the value.
\item Name of another variable the value is assigned to.
\end{enumerate}
  \end{minipage}}
\end{table}

Note that the acquired model predicts the usage patterns (and the
associated names) of just {\it one} variable.  This means that we need
to create a different probabilistic model for every single variable
that has different sets of usage patterns. We cannot use the same model
for all the variables in the program since it is contaminated with
their original names. We must exclude the hints of the original name
when we are trying to predict it.

The overall procedure of our system is as follows:

\begin{itemize}
\item Let $V$ as a set of all variables used in a target program.
\item For each $v_1 \in V$:
  \begin{enumerate}
  \item For all $v \in V$ $(v \ne v_1)$, obtain $P(v.\mathit{name} | v.\mathit{pattern})$.
    This is a model that predicts $v_1$.
  \item Find a name $n = \argmax P(n | v_1.\mathit{pattern})$.
  \item If $n = v_1$, the name is consistent with the model.
    Otherwise, suggest $n$ as a more appropriate name for $v_1$.
  \item A confidence score is calculated for each suggestion by adding
    the weights of the usage patterns:
    $\sum_{p \in v.\mathit{pattern}} \mathit{TF}(p) \times \mathit{IDF}(p) \times \exp(p.\mathit{dist})$ \footnote{$\mathit{TF}$ is the frequency of the pattern that is associated
      with the variable. $\mathit{IDF}$ is the inverse frequency of the pattern
      over all the outputs.
    } where
    $p$ is a usage pattern used for the suggestion and
    $p.\mathit{dist}$ is the distance between the pattern and its target variable.
  \end{enumerate}
\end{itemize}

In this paper, we use Na\"ive Bayes classifier for a probabilistic
model. In Na\"ive Bayes classification, learning a model is simply
counting the number of occurrences of $v.\mathit{name}$ and
$v.\mathit{pattern}$, and therefore it is easy to construct a
different model ($P$) for each variable \footnote{ We first learn a
model using all the usage patterns.  Then for each variable $v_1$ we
{\it unlearn} the patterns by subtracting the counts for
$v_1.\mathit{name}$ and $v_1.\mathit{pattern}$.  }.


\section{Experiments}


We applied to our method to the projects listed in Table \ref{tab:sources}.
We conducted three experiments for the following questions:

\begin{itemize} 
\item RQ1. Are usage patterns an effective representation for a variable usage?
\item RQ2. Did the system predict a correct variable name?
\item RQ3. Did the system provide convincing evidences to support the suggested alternatives?
\end{itemize}

Our results were evaluated by nine reviewers \footnote{ Three are the
  authors of this paper.  The other six are graduate students who has
  a basic experience of Java programming.}.  None of the reviewers
were familiar with the source code of a target project.  Reviewers
were asked not to talk about specific results during the experiments.
In the following subsections, we conducted experiments to
answer the above questions.

\begin{table}[tb]
  \caption{Source Code and Use-Flow Graph Sizes}
  \vspace*{-1em}
  \label{tab:sources}
  \centering
  \setlength \tabcolsep{4 pt}
  \begin{tabular}{|p{8em}|r|r|r|r|}
    \hline
    Project & kLOC & {\scriptsize Variables} & Nodes & Edges \\
    \hline
    ant 1.10.6 \newline {\scriptsize (build tool)}
    & 112k & 24k & 350k & 5,211k \\
    \hline
    antlr4 4.7.2 \newline {\scriptsize (parser generator)}
    & 31k & 7k & 74k & 1,103k \\
    \hline
    bcel 6.3.1 \newline {\scriptsize (Java bytecode library)}
    & 31k & 7k & 80k & 1,190k \\
    \hline
    compress 1.18 \newline {\scriptsize (compression library)}
    & 24k & 6k & 69k & 929k \\
    \hline
    jedit 5.5.0 \newline {\scriptsize (text editor)}
    & 115k & 22k & 294k & 6,106k \\
    \hline
    jhotdraw 5.3 \newline {\scriptsize (drawing tool)}
    & 80k & 17k & 235k & 2,351k \\
    \hline
    junit4 4.12 \newline {\scriptsize (unit testing)}
    & 9k & 2k & 21k & 280k \\
    \hline
    lucene 7.7.2 \newline {\scriptsize (document indexing)}
    & 109k & 30k & 414k & 7,146k \\
    \hline
    tomcat 8.5.43 \newline {\scriptsize (application server)}
    & 238k & 49k & 649k & 11,799k \\
    \hline
    weka 3.8 \newline {\scriptsize (machine learning)}
    & 324k & 59k & 943k & 13,224k \\
    \hline
    xerces 2.12.0 \newline {\scriptsize (XML parser)}
    & 114k & 22k & 314k & 7,017k \\
    \hline
    xz 1.18 \newline {\scriptsize (compression library)}
    & 7k & 2k & 23k & 299k \\
    \hline
  \end{tabular}
\end{table}

Our experimental setup was a standard desktop PC \footnote{Intel i5,
  1.8GHz, 32GB memory.} running Arch Linux. Extracting UFGs and
collecting usage patterns from source code took from a few minutes
to several hours, depending on the project size. Building a
probabilistic model and generating suggestions took several minutes.
All the tools and datasets that we used for this experiment
are publicly available \footnote{\tt https://github.com/euske/fgyama}.

\subsection{Variable Equivalence Test (RQ1)}

In the first experiment, we tested if two variables with similar usage
patterns have indeed a similar role.  This was done by collecting
pairs of variables whose usage patterns are similar to each other (the
similarity $>0.90$), and check if the two variables have a similar
name (role). The similarity is computed by taking the cosine distance
of two usage patterns as TF-IDF vectors, as in
\[
\mathit{Sim}(p_1, p_2) = \frac{V_1 \cdot V_2}{|V_1| |V_2|},
\hspace*{1em}
V_i = \sum \mathit{TF}(n_i) \times \mathit{IDF}(n_i).
\]
where $p_i$ and $n_i$ is a pattern and its nodes, respectively.

Then we presented the pairs of variables to the reviewers while hiding
the actual variable names by replacing them with ``{\tt xxx}''. The
reviewers were asked to look at each variable pair with its
surrounding code snippets and choose one of the following options:

\begin{enumerate}[label=(\alph*)] 
\item Variables {\it must} have the {\it same} name. ({\bf Must-Eq.})
\item Variables {\it can} have the {\it same} name. ({\bf Can-Eq.})
\item Variables {\it must} have a {\it different} name. ({\bf Must-Neq.})
\item Undecidable. ({\bf Unk.})
\end{enumerate}

The nine reviewers are presented with randomly selected five variable
pairs for 12 projects each.  Table \ref{tab:vareqs} shows the
responses. Out of 540 answers, 369 (68\%) was either {\bf Must-Eq.} or
{\bf Can-Eq.}. This suggests that usage patterns are a strong
indicator of the role of a variable.  The average cosine similarity of
{\bf Must-Eq.} or {\bf Can-Eq.} pairs was 0.980, whereas the
similarity of {\bf Must-Neq.} was 0.976.

\begin{table}
  \caption{Variable Equivalence Test}
  \vspace*{-1em}
  \label{tab:vareqs}
  \centering
  \setlength \tabcolsep{4 pt}
  \begin{tabular}{|l|r|r|r|r|r|}
    \hline
    Project & Must-Eq. & Can-Eq. & Must-Neq. & Unk. & Total \\
    \hline
    ant & 23 & 17 & 5 & 0 & 45 \\
    antlr4 & 13 & 22 & 9 & 1 & 45 \\
    bcel & 3 & 11 & 31 & 0 & 45 \\
    compress & 10 & 9 & 24 & 2 & 45 \\
    jedit & 7 & 25 & 9 & 4 & 45 \\
    jhotdraw & 15 & 17 & 12 & 1 & 45 \\
    junit4 & 14 & 21 & 10 & 0 & 45 \\
    lucene & 18 & 21 & 6 & 0 & 45 \\
    tomcat & 11 & 27 & 6 & 1 & 45 \\
    weka & 4 & 28 & 12 & 1 & 45 \\
    xerces & 12 & 22 & 9 & 2 & 45 \\
    xz & 4 & 15 & 26 & 0 & 45 \\
    \hline
    Total & 134 & 235 & 159 & 12 & 540 \\
    \hline
  \end{tabular}
\end{table}

\subsection{Name Suggestion Test (RQ2)}

\label{subsec:rq2}

In the second experiment, we tested if our proposed method can
actually check the usage of variable names and suggest a better
alternative for those which are found inconsistent with the other
parts of the program. This experiment is twofold; first, we presented
candidates of a variable name to the reviewers and let them choose the
best name among them, where one of the candidates is produced by our
system. We also manually created a patch for correcting some of the
prominent suggestions by our system and sent it to the original
developers of the projects.

\subsubsection{Evaluating System Outputs by Reviewers}

In the first part of the experiment, we presented code snippets to the
nine reviewers. For each question, a reviewer is presented with one
code snippet with one variable highlighted, and another snippet with a
different variable whose usage is similar to the first one, but whose
name is hidden.  The reviewers were asked to compare the two snippets
and infer an appropriate variable name for the hidden one. They can
choose from the following candidates, or choose undecidable ({\bf
  Unk.}):

\begin{enumerate}[label=(\alph*)] 
\item A name suggested by our system. ({\bf System})
\item A name suggested by a baseline system. ({\bf Base.})
\item The original name (chosen by the developer). ({\bf Orig.})
\end{enumerate}

\begin{figure}[tb]
\centering
\fbox{\includegraphics[width=0.7\linewidth]{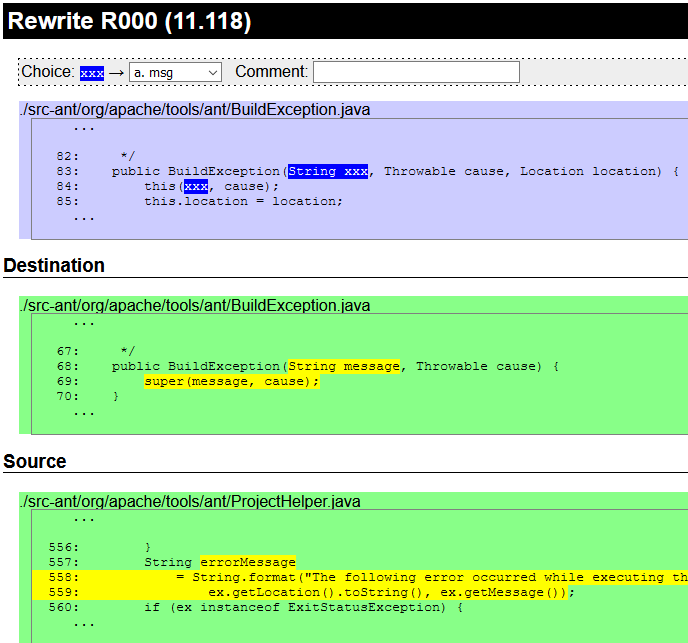}}
\caption{Evaluation Tool Screenshot}
\label{fig:evaltool}
\end{figure}

The baseline system here was only to suggest the most common name for
each data type. \footnote{For example, an {\tt int} variable is always
considered to be named {\tt i}.} The system produces a number of
suggestions for each project.  We ranked them by its confidence score
and present the top 10 suggestions to the reviewers.
The total number of generated suggestions (including ones that were not reviewed) and
its score distribution are shown in \ref{tab:scoredist}.
The screenshot of the evaluation tool is shown in Fig. \ref{fig:evaltool}.

Each reviewer answers 120 questions in total.  Table \ref{tab:cornames}
shows the reviewers' responses. Out of 1080 (= $9 \times 120$)
answers, 416 (39\%) was {\bf System}.  This means that for about 40\%
of the cases, our system can discover inconsistent variable names and
suggest alternatives which are considered better than the ones from
the original developers.

\begin{table}
  \caption{Generated Suggestions and Confidence Score Distribution \newline (by Percentile)}
  \vspace*{-1em}
  \label{tab:scoredist}
  \centering
  \setlength \tabcolsep{1 pt}
  \begin{tabular}{|l|r|r|r|r|r|r|r|r|r|r|r|}
    \hline
    Project & 10\% & 20\% & 30\% & 40\% & 50\% & 60\% & 70\% & 80\% & 90\% & 100\% & Total \\
    \hline
    ant & 2 & 15 & 61 & 144 & 302 & 637 & 1031 & 1985 & 3061 & 5784 & 13022 \\
    antlr4 & 7 & 9 & 41 & 74 & 144 & 217 & 344 & 453 & 658 & 965 & 2912 \\
    bcel & 3 & 5 & 12 & 55 & 103 & 203 & 294 & 501 & 644 & 1042 & 2862 \\
    compress & 4 & 6 & 16 & 45 & 98 & 214 & 276 & 504 & 822 & 1012 & 2997 \\
    jedit & 7 & 14 & 58 & 199 & 345 & 606 & 965 & 1621 & 2349 & 3837 & 10001 \\
    jhotdraw & 12 & 22 & 68 & 150 & 251 & 441 & 648 & 895 & 1284 & 1695 & 5466 \\
    junit4 & 6 & 12 & 27 & 40 & 89 & 89 & 89 & 149 & 182 & 135 & 818 \\
    lucene & 9 & 18 & 48 & 137 & 292 & 567 & 945 & 1735 & 2913 & 4546 & 11210 \\
    tomcat & 13 & 42 & 112 & 264 & 544 & 1165 & 1729 & 3038 & 4897 & 8422 & 20226 \\
    weka & 9 & 22 & 63 & 200 & 407 & 970 & 2054 & 3468 & 6905 & 10672 & 24770 \\
    xerces & 5 & 27 & 49 & 131 & 289 & 435 & 756 & 1263 & 1924 & 3371 & 8250 \\
    xz & 3 & 9 & 8 & 14 & 24 & 33 & 74 & 86 & 157 & 111 & 519 \\
    \hline
  \end{tabular}
\end{table}

\begin{table}
  \caption{Name Suggestion Test}
  \vspace*{-1em}
  \label{tab:cornames}
  \centering
  \begin{tabular}{|l|r|r|r|r|r|}
    \hline
    Project & System & Base. & Orig. & Unk. & Total \\
    \hline
    ant & 39 & 3 & 39 & 9 & 90 \\
    antlr4 & 34 & 3 & 46 & 7 & 90 \\
    bcel & 48 & 10 & 28 & 4 & 90 \\
    compress & 53 & 0 & 35 & 2 & 90 \\
    jedit & 24 & 4 & 42 & 20 & 90 \\
    jhotdraw & 4 & 8 & 78 & 0 & 90 \\
    junit & 34 & 1 & 48 & 7 & 90 \\
    lucene & 40 & 3 & 38 & 9 & 90 \\
    tomcat & 34 & 4 & 44 & 8 & 90 \\
    weka & 29 & 1 & 44 & 16 & 90 \\
    xerces & 31 & 1 & 49 & 9 & 90 \\
    xz & 46 & 1 & 37 & 6 & 90 \\
    \hline
    Total & 416 & 39 & 528 & 97 & 1080 \\
    \hline
  \end{tabular}
\end{table}

Since there is no gold standard for a variable name, we calculated the
Fleiss' Kappa \cite{fleiss_71} for measuring the inter-reviewers
agreement. Fleiss' Kappa is commonly used for measuring agreement
between $N$ people where $N \ge 3$. In case of $N = 2$, Cohen's Kappa
is typically used. The Fleiss' Kappa for our experiment was $K = 0.45$
(moderate agreement).

For exploring different ways of generating usage patterns, we changed
some parameters for feature generation and measured its accuracy
against our best output. Table \ref{tab:diff_features} shows how
different parameters can affect the system performance.

\begin{table}
  \caption{Performance by Different Feature Generation Parameters}
  \vspace*{-1em}
  \label{tab:diff_features}
  \centering
  \setlength \tabcolsep{4 pt}
  \begin{tabular}{|r|r|r|r|r|r|}
    \hline
    {\scriptsize Methods} & {\scriptsize Interproc?} & {\scriptsize Name?} & {\scriptsize Type?} & {\scriptsize Length?} & {\scriptsize Correct \%} \\
    \hline
    5 & + & + & + & 5 & {\bf 39\%} \\
    5 & + & + & + & 3 & 14\% \\
    5 & + & + & + & 1 & 7\% \\
    5 & + & - & - & 5 & 6\% \\
    5 & + & + & - & 5 & 10\% \\
    5 & + & - & + & 5 & 8\% \\
    5 & - & + & + & 5 & 10\% \\
    1 & + & + & + & 5 & 16\% \\
    \hline
  \end{tabular}
  \begin{description}[labelwidth=5em]
  \item[Methods]
    Maximum number of virtual methods for each function call.
  \item[Interproc?]
    If a usage pattern spans multiple functions.
  \item [Name?]
    If a feature about variable names are included.
  \item [Type?]
    If a feature about variable types are included.
  \item [Length?]
    Maximum length of usage patterns (number of nodes).
  \item [Correct \%]
    Ratio that the system suggestion was chosen.
  \end{description}
\end{table}

\subsubsection{Sending Patches to Developers}

In the second part of the experiment, we manually created a
patch for correcting some of the prominent suggestions by the system.
The patches were sent to the original developers of the projects.
We have gotten 6 responses so far. Three projects adopted our patch,
two are still discussing it, and one is rejected because ``this is not
a high priority'' according to the developer.

\subsection{Evidence Persuasiveness Test (RQ3)}

Our system can provide the evidences for suggested variable names.  An
``evidence'' is a code snippet where the prominent usage patterns for
the target variable were obtained from. This way, a user can review
the system outputs and decide if they can accept its result.  Since
each usage pattern has a weight and is associated with its original
location, the system can retrieve top $N$ \footnote{In this
experiment, we set $N = 3$.} usage patterns and its originating source
code. In this experiment, the reviewers are asked to review a variable
name suggestion with its original name and decide if the suggestion
is sensible based on the accompanying evidences (snippets). The
reviewers are asked to choose the following options regarding the
evidences.

\begin{enumerate}[label=(\alph*)] 
\item It provides strong support for the name. ({\bf Strong})
\item It provides reasonable support for the name. ({\bf Weak})
\item It provides little support for the name. ({\bf Poor})
\item Undecidable. ({\bf Unk.})
\end{enumerate}

Table \ref{tab:evidences} shows the reviewers' responses.  Out of 540
answers, 162 (30\%) of them are considered as somewhat supportive to
the suggested names. The relationship between the reviewers' ratings
and the system-generated confidence score is shown in Table
\ref{tab:perfscore}.  It is observable that suggestions with a lower
confidence score tend to be considered as a poor evidence.

\begin{table}
  \caption{Evidence Persuasiveness Test}
  \vspace*{-1em}
  \label{tab:evidences}
  \centering
  \begin{tabular}{|l|r|r|r|r|r|}
    \hline
    Project & Strong & Weak & Poor & Unk. & Total \\
    \hline
    ant & 7 & 8 & 30 & 0 & 45 \\
    antlr4 & 8 & 8 & 29 & 0 & 45 \\
    bcel & 10 & 15 & 20 & 0 & 45 \\
    compress & 5 & 5 & 34 & 1 & 45 \\
    jedit & 6 & 6 & 33 & 0 & 45 \\
    jhotdraw & 4 & 5 & 36 & 0 & 45 \\
    junit4 & 2 & 7 & 36 & 0 & 45 \\
    lucene & 3 & 14 & 27 & 1 & 45 \\
    tomcat & 7 & 7 & 31 & 0 & 45 \\
    weka & 1 & 10 & 34 & 0 & 45 \\
    xerces & 6 & 10 & 28 & 1 & 45 \\
    xz & 1 & 7 & 35 & 2 & 45 \\
    \hline
    Total & 60 & 102 & 373 & 5 & 540 \\
    \hline
  \end{tabular}
\end{table}

\begin{table}
  \caption{Performance by Confidence Score}
  \vspace*{-1em}
  \label{tab:perfscore}
  \centering
  \begin{tabular}{|l|r|r|r|r|r|}
    \hline
    \% Score & Total & Strong + Weak & Poor \\
    \hline
    $> 80\%$ & 207 & 71 (34\%) & 133 (64\%) \\
    $> 60\%$ & 108 & 40 (37\%) & 66 (61\%) \\
    $> 40\%$ & 108 & 20 (19\%) & 88 (81\%) \\
    $> 20\%$ & 108 & 24 (22\%) & 84 (78\%) \\
    $> 0\%$ & 9 & 7 (78\%) & 2 (22\%) \\
    \hline
  \end{tabular}
\end{table}

\subsection{Anecdotal Examples}

Here are a couple of anecdotal results (suggestions) that our system
produced:

\begin{itemize} \footnotesize
\item Make the name more task oriented. \vspace*{-1em} \\
\begin{minipage}[t]{\linewidth}
\begin{lstlisting}[frame=single, basicstyle=\scriptsize\ttfamily, escapechar=@]
- public static short getNoOfOperands(
-     final int @{\underline{index}}@) {
-   return NO_OF_OPERANDS[@{\underline{index}}@];
+ public static short getNoOfOperands(
+     final int @{\underline{opcode}}@) {
+   return NO_OF_OPERANDS[@{\underline{opcode}}@];
\end{lstlisting}
\end{minipage}

\item Use the conventional abbreviation of the project. \vspace*{-1em} \\
\begin{minipage}[t]{\linewidth}
\begin{lstlisting}[frame=single, basicstyle=\scriptsize\ttfamily, escapechar=@]
- void errorWhileMapping( String @{\underline{s}}@ ) {
+ void errorWhileMapping( String @{\underline{msg}}@ ) {
\end{lstlisting}
\end{minipage} \\
\begin{minipage}[t]{\linewidth}
\begin{lstlisting}[frame=single, basicstyle=\scriptsize\ttfamily, escapechar=@]
- String @{\underline{pkgName}}@ = className.substring(...);
+ String @{\underline{packageName}}@ = className.substring(...);
\end{lstlisting}
\end{minipage}

\item Use a synonym which aligns better with the other parts of the code. \vspace*{-1em} \\
\begin{minipage}[t]{\linewidth}
\begin{lstlisting}[frame=single, basicstyle=\scriptsize\ttfamily, escapechar=@]
- ReferencePosition(int n, int @{\underline{pos}}@) {
+ ReferencePosition(int n, int @{\underline{offset}}@) {
\end{lstlisting}
\end{minipage}

\item Correct typo. \vspace*{-1em} \\
\begin{minipage}[t]{\linewidth}
\begin{lstlisting}[frame=single, basicstyle=\scriptsize\ttfamily, escapechar=@]
- void normalize(int @{\underline{normalizeOffset}}@) {
+ void normalize(int @{\underline{normalizationOffset}}@) {
\end{lstlisting}
\end{minipage}

\end{itemize}


\section{Discussions}


Our experiments showed that UFGs and usage patterns can be effectively
used for discerning the use of variables. The system does not rely on
any predefined knowledge other than the language syntax, and
can be applied to a realistic project with modest computational resource.

As for the relevance of the experiment, note that our reviewers were
not familiar with the codebase used for the evaluation. We made sure
that all the reviewers (including the authors) be not familiar with a
target project in advance, and they do not discuss a specific example
during the experiment. However, the reviewers had an advantage of
seeing the different parts of the code side-by-side and being
presented a direct evidence of inconsistency, whereas the original
developers worked on only one part of the code.  This way, we argue
that it is possible that the reviewers made a better decision for
variable naming than the original developers who are obviously more
knowledgeable about the code.

\subsection{Threats to Validity}

There are a couple of threats to internal validity of our
experiments. First, our evaluation is subjective; it is affected by
the number of reviewers and their programming knowledge. One could
argue that nine reviewers are not enough. However, our Fleiss' Kappa
$K = 0.45$, which is far from random, suggests that our reviewers had
some common standard about variable naming.  Another concern is the
fairness of the reviews.  To address the reviewers' bias, we
randomized the order of the system output that is presented to a
reviewer so that they cannot know which name was produced by the
system.  However, in the Variable Equivalence Test (RQ1) we used
variables that already have a certain similarity ($Sim > 0.90$). As a
result, this test was not completely blinded.

When generalizing our results to a wider use, the threats to its
external validity are the following: There are not enough projects
tested. The target language for now is limited to Java.  There is a
limitation of our UFG extraction program that cannot track dynamic
dispatch and variable aliasing, which could be problematic for
expanding our method to pointer-rich programming languages like C++. A
bigger concern is that our experiment only measured the accuracy
(precision) of the system output, but not the overall coverage
(recall). There are concerns related to the performance of the machine
learning algorithm. In this paper, we tried to focus on our overall
framework and keep the learning part lean. However, we expect that
better algorithms (such as Recurrent Neural Network or Conditional
Random Field) can produce a better result. One could use a better
metrics for comparing usage patterns than ours (Cosine similarity,
TF-IDF).



\section{Conclusion}

In this paper, we presented a novel way of testing the semantic
consistency of variable names. We presented a graph structure called
Use-Flow Graph which is intended to capture the intent of programmers
from source code. Our system learned a naming convention of variables
by collecting usage patterns that are represented as a path of UFG
nodes.  We built a probabilistic model to infer a variable name from
its usage patterns. We applied our method to 12 open-source projects
and evaluated its results with human reviewers. The experiment showed
that our system identified inconsistent variable names and suggest
better names reasonably well compared to human developers. We plan to
extend the concept of UFGs to apply other program comprehension tasks
in future.


\bibliographystyle{IEEEtran}
\bibliography{../euske}

\begin{thebibliography}{10}
\providecommand{\url}[1]{#1}
\csname url@samestyle\endcsname
\providecommand{\newblock}{\relax}
\providecommand{\bibinfo}[2]{#2}
\providecommand{\BIBentrySTDinterwordspacing}{\spaceskip=0pt\relax}
\providecommand{\BIBentryALTinterwordstretchfactor}{4}
\providecommand{\BIBentryALTinterwordspacing}{\spaceskip=\fontdimen2\font plus
\BIBentryALTinterwordstretchfactor\fontdimen3\font minus
  \fontdimen4\font\relax}
\providecommand{\BIBforeignlanguage}[2]{{%
\expandafter\ifx\csname l@#1\endcsname\relax
\typeout{** WARNING: IEEEtran.bst: No hyphenation pattern has been}%
\typeout{** loaded for the language `#1'. Using the pattern for}%
\typeout{** the default language instead.}%
\else
\language=\csname l@#1\endcsname
\fi
#2}}
\providecommand{\BIBdecl}{\relax}
\BIBdecl

\bibitem{ap_04}
A.~Editors, \emph{The Associated Press Stylebook and Briefing on Media
  Law}.\hskip 1em plus 0.5em minus 0.4em\relax Lorenz Press, 2004.

\bibitem{nyt_99}
A.~M. Siegal and W.~G. Connolly, \emph{The New York Times Manual of Style and
  Usage}.\hskip 1em plus 0.5em minus 0.4em\relax Three Rivers Press, 1999.

\bibitem{kernighan_99}
B.~W. Kernighan and R.~Pike, \emph{The Practice of Programming}.\hskip 1em plus
  0.5em minus 0.4em\relax Boston, MA, USA: Addison-Wesley Longman Publishing
  Co., Inc., 1999.

\bibitem{mcconnel_04}
S.~McConnell, \emph{Code Complete, Second Edition}.\hskip 1em plus 0.5em minus
  0.4em\relax Redmond, WA, USA: Microsoft Press, 2004.

\bibitem{lawrie_06b}
\BIBentryALTinterwordspacing
D.~Lawrie, C.~Morrell, H.~Feild, and D.~Binkley, ``What's in a name? a study of
  identifiers,'' in \emph{Proceedings of the 14th IEEE International Conference
  on Program Comprehension}, ser. ICPC '06.\hskip 1em plus 0.5em minus
  0.4em\relax Washington, DC, USA: IEEE Computer Society, 2006, pp. 3--12.
  [Online]. Available: \url{https://doi.org/10.1109/ICPC.2006.51}
\BIBentrySTDinterwordspacing

\bibitem{beniamini_17}
\BIBentryALTinterwordspacing
G.~Beniamini, S.~Gingichashvili, A.~K. Orbach, and D.~G. Feitelson,
  ``Meaningful identifier names: The case of single-letter variables,'' in
  \emph{Proceedings of the 25th International Conference on Program
  Comprehension}, ser. ICPC '17.\hskip 1em plus 0.5em minus 0.4em\relax
  Piscataway, NJ, USA: IEEE Press, 2017, pp. 45--54. [Online]. Available:
  \url{https://doi.org/10.1109/ICPC.2017.18}
\BIBentrySTDinterwordspacing

\bibitem{avidan_17}
\BIBentryALTinterwordspacing
E.~Avidan and D.~G. Feitelson, ``Effects of variable names on comprehension an
  empirical study,'' in \emph{Proceedings of the 25th International Conference
  on Program Comprehension}, ser. ICPC '17.\hskip 1em plus 0.5em minus
  0.4em\relax Piscataway, NJ, USA: IEEE Press, 2017, pp. 55--65. [Online].
  Available: \url{https://doi.org/10.1109/ICPC.2017.27}
\BIBentrySTDinterwordspacing

\bibitem{host_09}
\BIBentryALTinterwordspacing
E.~W. H\o{}st and B.~M. Ostvold, ``Debugging method names,'' in
  \emph{Proceedings of the 23rd European Conference on ECOOP 2009 ---
  Object-Oriented Programming}, ser. Genoa.\hskip 1em plus 0.5em minus
  0.4em\relax Berlin, Heidelberg: Springer-Verlag, 2009, pp. 294--317.
  [Online]. Available: \url{http://dx.doi.org/10.1007/978-3-642-03013-0_14}
\BIBentrySTDinterwordspacing

\bibitem{allamanis_15}
\BIBentryALTinterwordspacing
M.~Allamanis, E.~T. Barr, C.~Bird, and C.~Sutton, ``Suggesting accurate method
  and class names,'' in \emph{Proceedings of the 2015 10th Joint Meeting on
  Foundations of Software Engineering}, ser. ESEC/FSE 2015.\hskip 1em plus
  0.5em minus 0.4em\relax New York, NY, USA: ACM, 2015, pp. 38--49. [Online].
  Available: \url{http://doi.acm.org/10.1145/2786805.2786849}
\BIBentrySTDinterwordspacing

\bibitem{mikolov_13}
\BIBentryALTinterwordspacing
T.~Mikolov, I.~Sutskever, K.~Chen, G.~Corrado, and J.~Dean, ``Distributed
  representations of words and phrases and their compositionality,''
  \emph{CoRR}, vol. abs/1310.4546, 2013. [Online]. Available:
  \url{http://arxiv.org/abs/1310.4546}
\BIBentrySTDinterwordspacing

\bibitem{alon_19}
\BIBentryALTinterwordspacing
U.~Alon, M.~Zilberstein, O.~Levy, and E.~Yahav, ``Code2vec: Learning
  distributed representations of code,'' \emph{Proc. ACM Program. Lang.},
  vol.~3, no. POPL, pp. 40:1--40:29, Jan. 2019. [Online]. Available:
  \url{http://doi.acm.org/10.1145/3290353}
\BIBentrySTDinterwordspacing

\bibitem{raychev_15}
\BIBentryALTinterwordspacing
V.~Raychev, M.~Vechev, and A.~Krause, ``Predicting program properties from "big
  code",'' \emph{SIGPLAN Not.}, vol.~50, no.~1, pp. 111--124, Jan. 2015.
  [Online]. Available: \url{http://doi.acm.org/10.1145/2775051.2677009}
\BIBentrySTDinterwordspacing

\bibitem{liu_19}
\BIBentryALTinterwordspacing
K.~Liu, D.~Kim, T.~F. Bissyand\'{e}, T.~Kim, K.~Kim, A.~Koyuncu, S.~Kim, and
  Y.~L. Traon, ``Learning to spot and refactor inconsistent method names,'' in
  \emph{Proceedings of the 41st International Conference on Software
  Engineering}, ser. ICSE ’19.\hskip 1em plus 0.5em minus 0.4em\relax IEEE
  Press, 2019, p. 1–12. [Online]. Available:
  \url{https://doi.org/10.1109/ICSE.2019.00019}
\BIBentrySTDinterwordspacing

\bibitem{allamanis_14}
\BIBentryALTinterwordspacing
M.~Allamanis, E.~T. Barr, C.~Bird, and C.~Sutton, ``Learning natural coding
  conventions,'' in \emph{Proceedings of the 22nd ACM SIGSOFT International
  Symposium on Foundations of Software Engineering}, ser. FSE 2014.\hskip 1em
  plus 0.5em minus 0.4em\relax New York, NY, USA: Association for Computing
  Machinery, 2014, p. 281–293. [Online]. Available:
  \url{https://doi.org/10.1145/2635868.2635883}
\BIBentrySTDinterwordspacing

\bibitem{basili_96}
\BIBentryALTinterwordspacing
V.~R. Basili and S.~K. Abd-El-Hafiz, ``A method for documenting code
  components,'' \emph{J. Syst. Softw.}, vol.~34, no.~2, pp. 89--104, Aug. 1996.
  [Online]. Available: \url{http://dx.doi.org/10.1016/0164-1212(95)00070-4}
\BIBentrySTDinterwordspacing

\bibitem{lawrie_06a}
D.~{Lawrie}, H.~{Feild}, and D.~{Binkley}, ``Syntactic identifier conciseness
  and consistency,'' in \emph{2006 Sixth IEEE International Workshop on Source
  Code Analysis and Manipulation}, Sep. 2006, pp. 139--148.

\bibitem{deissenboeck_06}
\BIBentryALTinterwordspacing
F.~Deissenboeck and M.~Pizka, ``Concise and consistent naming,'' \emph{Software
  Quality Journal}, vol.~14, no.~3, pp. 261--282, Sep. 2006. [Online].
  Available: \url{http://dx.doi.org/10.1007/s11219-006-9219-1}
\BIBentrySTDinterwordspacing

\bibitem{engler_01}
\BIBentryALTinterwordspacing
D.~Engler, D.~Y. Chen, S.~Hallem, A.~Chou, and B.~Chelf, ``Bugs as deviant
  behavior: A general approach to inferring errors in systems code,''
  \emph{SIGOPS Oper. Syst. Rev.}, vol.~35, no.~5, p. 57–72, Oct. 2001.
  [Online]. Available: \url{https://doi.org/10.1145/502059.502041}
\BIBentrySTDinterwordspacing

\bibitem{miller_95}
\BIBentryALTinterwordspacing
G.~A. Miller, ``Wordnet: A lexical database for english,'' \emph{Commun. ACM},
  vol.~38, no.~11, pp. 39--41, Nov. 1995. [Online]. Available:
  \url{http://doi.acm.org/10.1145/219717.219748}
\BIBentrySTDinterwordspacing

\bibitem{reps_97}
\BIBentryALTinterwordspacing
T.~Reps, ``Program analysis via graph reachability,'' in \emph{Proceedings of
  the 1997 International Symposium on Logic Programming}, ser. ILPS '97.\hskip
  1em plus 0.5em minus 0.4em\relax Cambridge, MA, USA: MIT Press, 1997, pp.
  5--19. [Online]. Available:
  \url{http://dl.acm.org/citation.cfm?id=271338.271343}
\BIBentrySTDinterwordspacing

\bibitem{fleiss_71}
\BIBentryALTinterwordspacing
J.~L. Fleiss, ``Measuring nominal scale agreement among many raters,''
  \emph{Psychological Bulletin}, vol.~76, no.~5, p. 378–382, 1971. [Online].
  Available: \url{https://doi.org/10.1037/h0031619}
\BIBentrySTDinterwordspacing

\end{thebibliography}

\end{document}